\documentstyle[aps,pre,multicol,epsf]{revtex}
\begin{document}
\def\OP {\tensor P}
\def\B.#1{{\bbox{#1}}}
\def\C.#1{{\cal{#1}}}
\def\o {\omega}
\def\BE{\begin{equation}}
\def\BEA{\begin{eqnarray}}
\def\EE{\end{equation}}
\def\EEA{\end{eqnarray}}
\def\a{\alpha}
\def\b{\beta}
\def\g{\gamma}
\def\d{\delta}
\def\z{\zeta}
\def\k{\kappa}
\def\nn{\nonumber}
\def\emv{\sf}
\renewcommand{\thesection}{\arabic{section}}
\title{Temporal surrogates of spatial turbulent statistics:
\\the Taylor    hypothesis revisited}
\author {Victor S. L'vov, Anna Pomyalov and Itamar Procaccia}
\address{Department of~~Chemical Physics, The Weizmann
Institute of Science, Rehovot 76100, Israel} \maketitle

\begin{abstract}
The Taylor hypothesis which allows surrogating spatial measurements
requiring many experimental probes by time series from one or two
probes is examined on the basis of a simple analytic model of
turbulent statistics. The main points are as follows: (i) The
Taylor hypothesis introduces {\em systematic} errors in the evaluation
of scaling exponents. 
(ii) When the mean wind $\overline{V}_0$
is not infinitely larger than the root-mean-square longitudinal turbulent
 fluctuations $v_{_{\rm T}}$, the  effective Taylor advection velocity
$V_{\rm ad}$ should take the latter into account.
 (iii) When two or more probes are employed 
 the application of the Taylor hypothesis and the optimal  choice of
the effective advecting  wind $V_{\rm ad}$  need
 extra care. We present practical considerations for
minimizing the errors incurred in experiments using
one or two probes. (iv) Analysis of the Taylor hypothesis when different probes
experience different mean winds is offered.
\end{abstract}
\begin{multicols}{2}
\section{Introduction}
Decades of research on the statistical aspects of thermodynamic
turbulence are based on the Taylor hypothesis \cite{38Tay}, which
asserts that the fluctuating velocity field measured by a given probe
as a function of time, $\B.u(t)$ is the same as the velocity
$\B.u(R/\overline{V\! _0 })$ where $\overline{V\! _0 }$ is
the mean velocity and $R$  is the distance to
a position ``upstream'' where the velocity is measured at $t=0$. Sixty
years after its introduction by Taylor, this time honored hypothesis
remains the only really convenient way to measure experimentally
turbulent velocity fluctuations. New techniques were introduced in
recent years, but so far did not make a lasting mark on the field. On
the other hand theoretical considerations of the anomalous nature of
the statistics of turbulence have made higher and higher demands on
the accuracy of experimental measurements, with finer details being
asked by experimentalists and theorists alike. In light of these
demands it seems necessary to revisit the Taylor hypothesis at this
point to assess its consequences regarding the accuracy of
measurements of scaling exponents in turbulent media.

Our own motivation to study the consequences of the Taylor hypothesis
stems from attempts to develop deeper understanding of the effects of
anisotropy on turbulent statistics \cite{98ADKLPS,99ALPa}. In the
context of this program it turned out that the interpretation of
experimental signals in turbulent systems with shear poses delicate
issues that call for careful considerations. In order to expose
anisotropic features one needs to analyze data pertaining to at least
two probes. In the case of shear each probe may experience a different
mean velocity, and velocity differences between such two probes (which
are computed using Taylor surrogates) mix
spatial and temporal dependencies. The considerations taken to clarify
such issues are assisted by the analysis of a simple model of turbulent
advection, which sheds light on how to treat systems with shear, but
also can be used to improve the understanding of the Taylor hypothesis
in systems that are homogeneous and isotropic. It seems therefore
worthwhile to present the model and its consequence for the benefit of
the general turbulence community which may find it useful for more
than one application.

The Taylor hypothesis was studied carefully in the 50's
\cite{48Wei,53Ogu,56Gif,75MY},
and continues to be the subject of scrutiny to this day
\cite{91Lvo,94PL,98SD}.
Some of the inherent limitations implied by the
Taylor hypothesis were pointed out in these studies. Our purpose in
this paper is to offer rational choices to minimize the systematic
errors that are entailed in the standard experimental procedures. To
this aim we need to study the systematic errors, something that can be
done only by comparing spatial statistics to temporal statistics. Not
being able to do this directly on the basis of the Navier-Stokes
equations, we offer a model of turbulent fluctuations advected by a
``wind'' of desired properties, be them homogeneous or not. The model
allows us to compute explicitly correlation functions or structure
functions that depend on space and time. We can then compare the
temporal objects (for fixed spatial positions) with simultaneous
objects that depend on varying scales. Having full control on the
properties of the wind we can analyze the relative importance of the
mean wind versus the rms fluctuations and the consequences of
inhomogeneities.

In Sect. 2 we present the issue, introduce the statistical objects under
study, and explain the model that is analyzed in the rest of this
paper. The model employs an advecting velocity field $\B.V$ and an independent
fluctuating field $\B.u$ which is advected without affecting its statistical
properties. The latter are chosen to mimic those of Kolmogorov turbulence.
The most important
property that affects the accuracy of the Taylor surrogate is the effective
decay time of fluctuations of scale $R$. The ratio of the sweeping time across
a scale $R$ and this decay time determines the applicability of the Taylor
hypothesis. This is made clear in Sect. 2. In Sect. 3
we explore the consequences of the Taylor hypothesis in the case of one probe
measurements. We find that the Taylor method introduces systematic errors
in the estimated exponents of the 2nd order structure function. The reason for
this error is simply that the Taylor method improves for small scales, where
the decay time is always much longer than the sweeping time. Accordingly, there
is a systematic improvement of the estimate via surrogates as the relevant
length scale decreases. This appears as an apparent ``exponent" in log-log
plots. Nevertheless, we argue that the systematic errors for the isotropic
part of the 2nd order structure function are quite
small for realistic choices of the parameters, of the order of 0.01 in the
measured
exponents. In the same Section we discuss the relative contribution of the
mean wind and the rms fluctuations to the ``effective" advecting wind
$V_{\rm ad}$ employed in the
Taylor hypothesis. We find that  the method works even in the absence
of mean wind (which has been noticed before, for example in turbulent
convection \cite{91Lvo} and in a swirling flow \cite{94PL}). In general both
contribute to the effective wind, with a parameter  of relative importance
[denoted $b$ below, see Eq.~(\ref{defV*})]. We find that the optimal value
of $b$ is larger than anticipated.

In Section~4 we solve the model in the case of linear shear. The first
question analyzed is what is the effective wind that should be taken in
surrogating
data that stem from two probes that experience different mean winds.
 We show
that for linear shear the answer is simple, {\em i.e.} the mean of the mean
winds
of the two probes. Next we solved the model, and found the corrections
to the structure functions due to the existence of the shear. In the language
of Ref. \cite{99ALPb} this is a $j=2$ anisotropic contribution where
$j$ refers to the index of the irreducible representation of the SO(3)
symmetry group. The scaling exponent associated with this contribution
is 4/3 in the K41 framework, in agreement with
measurements and earlier theoretical
considerations\cite{73Les,99KLPS} . Lastly we assessed the performance
of the Taylor method for
this contribution and concluded that it is significantly worse than in the
isotropic
counterpart. The typical errors in estimating the exponent can be as high
as 0.1.
Section~5 offers a summary and a discussion.  In particular we present
arguments as to which aspects  of our conclusions are relatively model
independent.
\section{The Model}
\subsection{Preliminaries}
In statistical turbulence one is interested in the statistical
properties of the turbulent velocity field $\B.u(\B.r,t)$ where
$(\B.r,t)$ is a space-time point in the laboratory frame (so called
``Eulerian'' velocity).  In this paper we will focus on the properties
of the second order space-time correlation function of velocity
differences:
\begin{eqnarray}
F^{\alpha\beta}(\B.R,t)&\equiv&
\Big\langle\big[u^\alpha(0,t_0)-u^\alpha(\B.R,t_0+t)\big]
\nonumber\\&\times&\big[u^\beta(0,t_0)
-u^\beta(\B.R,t_0+t)\big]\Big\rangle \
, \label{Fab}
\end{eqnarray}
where angular brackets denote averaging with respect to $t_0$. In this
definition and throughout the paper we assume that the turbulence is
{\em stationary} in the sense that the statistical ensemble is time
independent. We do not assume space homogeneity or
isotropy. For $t=0$ the correlation function $F^{\alpha\beta}(\B.R,t)$
turns into the commonly used second-order structure function
$S^{\alpha\beta}(\B.R)$:
\begin{equation}
S^{\alpha\beta}(\B.R)\equiv F^{\alpha\beta}(\B.R,t=0) \ . \label{Sab}
\end{equation}
For $\B.R=0$ we have the time-dependent object which is usually measured in
single probe experiments:
\begin{equation}
T^{\alpha\beta}(t)\equiv F^{\alpha\beta}(\B.R=0,t) \ . \label{Tab}
\end{equation}
The Taylor hypothesis is based on the idea that when the mean wind
$\overline{\B.V\! _0 }$ is very high, the turbulent field is advected by a
given probe
as if frozen, having hardly any time to relax while being recorded by
the probe.  Disregarding the relaxation of turbulent eddies of size
$R$, the hypothesis implies that
\begin{equation}
S^{\alpha\beta}(\B.R=t\overline{\B.V\! _0 }) = T^{\alpha\beta}(t) \ ,
\label{Taylor}
\end{equation}
Obviously, the validity of this hypothesis depends on the ratio of two
times scales. The first is the advection time $R/\overline{V\! _0 }$ which
takes to
translate structures of size $R$ by the probe. The second is the
life-time $\tau(R)$ which describes the typical decay time of turbulent
structures of size $R$. In the limit $R/[ \overline{V\! _0 }\tau(R)]\to 0$
the Taylor
hypothesis becomes valid.  The typical time scale $\tau(R)$ is
inherent to the dynamics of turbulent flows, and is quite independent
of the mean wind which can be eliminated by changing the coordinates
to a co-moving frame. Up to a factor of order unity the life-time can
be estimated as the turn-over time $R/\sqrt{S(R)}$ where $S(R)\equiv
S_{\alpha\alpha}(R)$.  With this estimate the Taylor hypothesis is
expected to be valid when $\sqrt{S(R)}/\overline{V\! _0 } \to 0$. In the
sequel we
denote the ratio of these two time scales by $z(R)$. Clearly, in
turbulence $z(R)$ increases with $R$, and for $R$ of the order of the
outer scale of turbulence it is largest. It is thus sufficient to have
very small $z(L)$ to ensure the validity of the Taylor hypothesis for
all $r<L$.

In typical experimental conditions, like atmospheric turbulence,
$z(L)$ is of the order of 0.2-0.5 \cite{98SD,94PO}. (Note that in most
experimental papers only the longitudinal component of the structure
function is available; in isotropic turbulence this is smaller than
$S(R)$ by a factor of about 3. Accordingly, the Taylor hypothesis needs
careful scrutiny.  Moreover, almost all experiments are forced by
anisotropic and inhomogeneous agents, and the ``mean'' velocity
depends on the position. When more than one probe is used one needs to
decide how to choose $\overline{V\! _0 }$ in Eq. (\ref{Taylor}).  To allow
us to answer
such questions rationally we study the following model.
\subsection{Basic Model}
\subsubsection{Equation of Motion }
Consider  a model turbulent velocity field $\B.u(\B.r,t)$ which in ($\B.k,
\omega$)-representation  is defined as
\begin{equation}
\tilde\B.u(\B.k,\omega)=\int d\B.r  \exp[-i(\B.r\cdot\B.k+\omega
t)]\B.u(\B.r,t) \ . \label{Fourier}
\end{equation}
We propose the following model dynamics for $\tilde\B.u(\B.k,\omega)$:
\begin{eqnarray}\label{model}
&&\Big[\omega+\B.k\cdot
\B.V\!_0+i\gamma(k)\Big]\tilde u^\alpha (\B.k,\omega)
+\int{d\B.k'd\B
.k''\over 8\pi^3} \\ \nonumber
&&\times
\Gamma^{\alpha\beta\gamma}_{\B.k} V^\beta_{\rm s}(\B.k')\tilde
u^\gamma(\B.k'',\omega)
\delta(\B.k-\B.k'-\B.k'')=\tilde f^\alpha(\B.k,\omega) \ ,
\\
&&\quad i\B.k\cdot \tilde\B.u(\B.k,\omega)=0\,, \label{incom}
\end{eqnarray}
where $\Gamma_{\B.k}^{\alpha\beta\gamma}$ is the exact nonlinear vertex that
stems from the Navier-Stokes equations:
\begin{equation}\label{vertex}
\Gamma_{\B.k}^{\alpha\beta\gamma}
=k_\beta P^{\alpha\gamma}(\B.k)
+k_\gamma P^{\alpha\beta} (\B.k)
 \ .
\end{equation}
Here $P^{\alpha\beta} (\B.k)$ is the transverse projection operator
\begin{equation}\label{tran}
P^{\alpha\beta} (\B.k)=\delta_{\alpha\beta}
-\frac{k_\alpha k_\beta}{k^2}\,,
\end{equation}
 and $\delta_{\alpha\beta}$ is the Kronecker symbol. This dynamics
 represents ``passive vector advection'' in which the ``turbulent''
 field $\tilde\B.u(\B.k,\omega)$ is advected by a statistically
 independent stationary field $\B.V(\B.k)$. In its turn, the wind
 $\B.V(\B.k)$ consist of homogeneous $\B.V\!_0$ and space dependent
 $\B.V_{\rm s} (\B.k)$ parts:
\begin{equation}\label{wind1}
\B.V(\B.k)=(2\pi)^3\delta(\B.k)\B.V\!_0
+\B.V_{\rm s} (\B.k)\ .
\end{equation}
The homogeneous part $\B.V\!_0$ appears in Eq.~(\ref{model}) as a
Doppler shift to $\omega$. The inverse decay time $\gamma(k)$
represents the eddy-viscosity which mimics the effects of the
nonlinear terms in Navier-Stokes dynamics on the energy loss from a
given wavenumber. The forcing term $\B.f(\B.k,\o)$ represents that
energy gain.

\subsubsection{Statistical description}
\noindent
$\bullet$ {\em Correlations in $(\B.k,\omega)$- and
$(\B.k,t)$-representation}: Introduce the correlation function of the
velocity field $\tilde\B.u(\B.k,\omega)$ as follows:
\begin{equation}
\langle \tilde u^\alpha(\B.k,\omega) \tilde
u^{*\beta}(\B.k',\omega')\rangle \equiv
2\pi\delta(\o-\o')\tilde \Phi^{\alpha\beta}
(\B.k,\B.k',\o) \ . \label{Phi}
\end{equation}
For space-homogeneous ensembles (in our case, in the absence of a
shear) $\tilde
\Phi^{\alpha\beta} (\B.k,\B.k',\o)$ is diagonal in $\B.k$:
\begin{equation}\tilde \Phi^{\alpha\beta}(\B.k,\B.k',\o) =
(2\pi)^3 \delta(\B.k-\B.k')\tilde \Phi^{\alpha\beta}(\B.k,\o)\ .
\label{Phi-d}
\end{equation}
Note that in order to avoid the proliferation of symbols we used
the same notation for the two functions
$\tilde\Phi^{\alpha\beta} (\B.k,\B.k',\o)$ and
$\tilde\Phi^{\alpha\beta}(\B.k,\o)$. The
same two functions in $\B.k,t$ representations are distinguished by a
``hat" symbol:
\begin{eqnarray}\label{Phi-t}
\hat \Phi^{\alpha\beta}(\B.k,\B.k',t) &
=&\int \frac{d\o}{2\pi}\tilde\Phi^{\alpha\beta}
(\B.k,\B.k',\o)\exp (i\o t)\,, \\ \nonumber
\hat \Phi^{\alpha\beta}(\B.k,t) &
=&\int \frac{d\o}{2\pi}\tilde\Phi^{\alpha\beta}
 (\B.k,\o)\exp (i\o t)\ .
\end{eqnarray}
The time independent functions $\hat
\Phi^{\alpha\beta}(\B.k,\B.k',t=0)$ and $\hat
\Phi^{\alpha\beta}(\B.k,t=0)$ will remain undecorated:
\begin{eqnarray}\label{Phi-0}
\Phi^{\alpha\beta}(\B.k,\B.k')\equiv
\hat \Phi^{\alpha\beta}(\B.k,\B.k',0)\,, \quad
\Phi^{\alpha\beta}(\B.k)\equiv
\hat \Phi^{\alpha\beta}(\B.k,0)\ .
\end{eqnarray}
\vskip 0.5cm
\noindent
$\bullet$ {\em Correlation functions in
$(\B.r,t)$-representation}: Introduce correlation functions of the
velocity filed $\B.u(\B.r,t)$ as follows:
\begin{equation}
\langle u^\alpha(\B.r,t)
u^{\beta}(\B.r',t')\rangle \equiv
\C.F^{\alpha\beta}
(\B.r,\B.r',t-t') \,, \label{Frrt}
\end{equation}
where stationarity in time is assumed. In space homogeneous ensembles
$\C.F^{\alpha\beta} (\B.r,\B.r',t) $ depends
on the difference $\B.R=\B.r-\B.r'$ only. We will again use an economic
notation and employ the symbol
$\C.F^{\alpha\beta}$ also for the space homogeneous case:
\begin{equation}
\C.F^{\alpha\beta}
(\B.r,\B.r',t) \Longrightarrow \C.F^{\alpha\beta}
(\B.R,t) \ .
 \label{Frt}
\end{equation}
These two functions are related to the corresponding correlation functions
in $\B.k,t$-representation by
\begin{eqnarray}\label{Frt-a}
\C.F^{\alpha\beta}(\B.r,\B.r',t) &
=&\int \!\! \frac{d\B.kd\B.k'}{(2\pi)^6}\hat \Phi^{\alpha\beta}
(\B.k,\B.k',t)\\ \nn
&& \times   \exp \!\big[i(\B.k\cdot\B.r-\B.k'\cdot\B.r')\big], \\
\C.F^{\alpha\beta}(\B.R,t) &
=&\int \!\! \frac{d\B.k}{(2\pi)^3}\hat \Phi^{\alpha\beta}
(\B.k,t)\exp (i\B.k\cdot\B.R) \ . \label{Frt-b}
\end{eqnarray}
On the other hand the function $F^{\alpha\beta}$ of Eq. (\ref{Fab})
is computed as
\begin{equation}
F^{\alpha\beta}(\B.R,t)=2\int \frac{d\B.k}{(2\pi)^3}
\hat \Phi^{\alpha\beta}
(\B.k,t)\left[1-\exp (i\B.k\cdot\B.R)\right] \ . \label{compF}
\end{equation}
\subsubsection{Choice of parameters in the model}
\noindent
$\bullet$ {\em The advecting wind}: In our thinking we are inspired by
experiments in the atmospheric boundary layer in which the advecting
wind may be considered as consisting of three parts. The first
component can be taken as a space-time independent mean wind
$\overline{\B.V\! _0 }$ which is constant for our ensemble. The second
component is a space time independent part which is constant on the
time scale of a typical experiment (minutes), but it changes from one
experimental realization in the ensemble to another.  We denote it as
$\B.V\!_{_{\rm T}}$. We will assume that it fluctuates randomly
between different experimental realizations of the ensemble. The third
part is an explicitly space dependent part of the mean wind denoted
as above $\B.V\!_{\rm s}(\B.r)$.  Note that again we avoid
proliferating the symbols, and we use the same symbol $\B.V_{\rm s}$
in $\B.k$ and $\B.r$ representation.  Accordingly we can write
\begin{equation}\label{wind2}
\B.V\!_0=   \overline{\B.V\! _0 }+\B.V\!_{_{\rm
T}}\,, \quad\overline{\B.V\!_{_{\rm T}}}=0\ .
\end{equation}
Since $\B.V\!_{_{\rm T}}$ is considered as a random variable we need
to specify its probability distribution function. This is denoted
${\cal P}(\B.V\!_{_{\rm T}})$, and overlines as in
Eq.~(\ref{wind2}) denote averages with respect to this
distribution.  We will solve the correlation functions $\tilde
\B.\Phi(\B.k,\o)$ for each realization of $\B.V\!_0$ and average the
result with respect to ${\cal P}(\B.V\!_{_{\rm T}})$. The amplitude of
the mean-square fluctuations of $\B.V\!_{_{\rm T}}$ are chosen such that
\begin{equation}
\overline {V\!_{_{\rm T}}^2}=3v\!_{_{\rm T}}^2 \,,\label{fixUv}
\end{equation}
where $v\!_{_{\rm T}}^2 $ is a mean-square fluctuation of the
longitudinal turbulent velocity.

The inhomogeneous part of the wind will not be random. To simplify the
analytical calculations the space dependent $\B.V_{\rm s}(\B.r)$ is
chosen as a sinusoidal profile
\begin{equation}
\label{prof1}
\B.V_{\rm s}(\B.r)=\B.n \,V_{\rm s}\sin(\B.q\cdot \B.r)\,, \quad
\B.q= q\B.m\,,
\end{equation}
where $\B.m$ and $\B.n$ are unit vectors in the vertical and horizontal
directions respectively. The horizontal direction is the direction of
the mean wind: $\overline{\B.V\! _0 }=\B.n \,\overline{V\! _0 }$.  In
$\B.k$ representation Eq.~(\ref{prof1}) reads:
\begin{equation}
\B.V_{\rm s} (\B.k) ={(2\pi)^3\over 2}V_{\rm s}\B.n
[\delta(\B.k-\B.q)-
\delta(\B.k+\B.q)] \ . \label{choiceV}
\end{equation}
Note that the sinusoidal profile~(\ref{prof1}) has nothing to do with
the logarithmic profile in real boundary layers. For small $\B.q$ it
mimics locally a linear shear.
\vskip0.5cm
\noindent
$\bullet$ {\em The life time of eddies}: A good model for
$\gamma(k)$ in Eq. (\ref{model}) is provided by the Kolmogorov 41
model of turbulence in
which the life time $1/\gamma(k)$ is defined as the turn-over time up
to an unknown dimensionless (universal) factor $C$:
\begin{equation}
\gamma(k)=C \frac{v_{_{\rm T}}}{L}(kL)^{2/3} \ . \label{defgam}
\end{equation}
Here $L$ is the integral scale of turbulence and $v_{_{\rm T}}^2$ is
the mean square longitudinal velocity which in isotropic conditions
equals
\begin{equation}
v_{_{\rm T}}^2 = \frac{1}{3}\langle |\B.u (\B.r,t)|
^2\rangle \ . \label{vT2}
\end{equation}

\vskip0.5cm
\noindent
$\bullet$ {\em The forcing term $\B.f(\B.k,\o)$}: In this paper we are
interested in second order turbulent statistics. Therefore it is
sufficient to model $\B.f(\B.k,\o)$ as Gaussian white noise:
\begin{eqnarray}\label{randomf}
&& \langle \tilde f^\alpha(\B.k,\o) \tilde
f^{*\beta}(\B.k',\o')\rangle \\ \nonumber
&=&(2\pi)^4 \delta(\o-\o')\delta(\B.k-\B.k')
D^{\alpha\beta}(\B.k)\ .
\end{eqnarray}
Since our model is linear in the turbulent velocity $\tilde \B.u$, there
is a simple relation between the intensity of the noise
$D^{\alpha\beta}(\B.k)$ and the simultaneous correlation
function of the turbulent velocity $\Phi_0^{\alpha\beta}(\B.k)$, where
the subscript~``$~_0$~`` denotes the absence of the shear flow. The relation
is (and cf. Eq. (\ref{Phi-k}) below):
\begin{equation}\label{fdt}
 D^{\alpha\beta}(\B.k)=2\gamma(k)\,\Phi_0^{\alpha\beta}(\B.k)
\ .
\end{equation}
The tensorial structure of
$\Phi_0^{\alpha\beta}(\B.k)$ is determined by the incompressibility condition
\begin{equation}\label{Phi0}
 \Phi_0^{\alpha\beta} (\B.k) =P^{\alpha\beta}(\B.k)\Phi_0(k)\ ,
\end{equation}
and what remains is to select the scalar function $\Phi_0(k)$. To do
this we refer again to the K41 model and choose
\begin{equation}
\Phi_0(\B.k)={\phi\over [(kL)^2+1]^{11/6}} \,,\label{choice}
\end{equation}
with some amplitude $\phi$. In the inertial interval, {\em i.e.}
for $k L\gg 1$. Eq. (\ref{choice}) agrees with the standard Kolmogorov
scaling, $\Phi_0(\B.k)\propto k^{-11/3}$. The form of
Eq. (\ref{choice}) is not unique, and other forms exhibiting different
crossovers between power law scaling and saturation are equally
acceptable. For example instead of (\ref{choice}) we may also choose
\begin{equation}
\Phi_0(\B.k)={\phi\over (kL)^{11/3}+1} \ . \label{choice2}
\end{equation}
We will show below
that our conclusions are only weakly affected by the precise choice of
crossover
behavior. This completes the set up of the model.

\section{Solutions of the model without shear}
\subsection{Homogeneous advection}
Firstly we analyze the situation without shear, $V_{\rm s}=0$.
The resulting velocity field $\tilde\B.u_0(\B.k,\o)$ and all the other
objects will be denoted by a subscript~``$~_0~$'' to remind us that $V_{\rm
s}=0$.
In this case the integral in Eq.~(\ref{model}) vanishes and the solution for
$\tilde\B.u_0(\B.k,\o)$ immediately follows:
\begin{eqnarray}\label{u0}
\tilde\B.u_0(\B.k,\o)&=&G_0 (\B.k,\o) \tilde\B.f(\B.k,\o)
  \\ \label{green}
G_0 (\B.k,\o)&\equiv& \frac{1}{\o+\B.k\cdot
\B.V\!_0+i\gamma(k)}  \,,
\end{eqnarray}
One sees that the effect of the space homogeneous
part of the advecting velocity field amounts to a Doppler shift
only. Using definitions~(\ref{Phi}), (\ref{Phi-d}) and (\ref{randomf})
one has:
\begin{equation}\label{Phi-ko}
\tilde \Phi^{\alpha\beta}_0(\B.k,\o)=D^{\alpha\beta}(\B.k)|
G_0 (\B.k,\o)|^2 \,,
\end{equation}
The equation for the  simultaneous correlation function follows from
(\ref{Phi-t}):
\begin{equation}\label{Phi-k}
\Phi^{\alpha\beta}_0(\B.k)=\int \frac{d\,\o}{2\pi}
\tilde \Phi^{\alpha\beta}_0(\B.k,\o)=\frac{D^{\alpha\beta}(\B.k)}
{2\gamma(k)}\ .
\end{equation}
This is consistent with Eq. (\ref{fdt}).
The correlation
function in ($\B.k,t$)-representation is computed straightforwardly,
\begin{eqnarray}
\hat \Phi_0^{\alpha\beta}(\B.k,t)&
=&\int{d\o\over 2\pi}\tilde
\Phi_0^{\alpha\beta}(\B.k,\o)\exp(i\o\tau)
\nonumber\\&=&
\Phi_0^{\alpha\beta}(\B.k) \exp[i\B.k
\cdot \B.V\!_0 t-\gamma(k) t] \ .
\label{solPhi}
\end{eqnarray}
At this point we recall that $\B.V\!_0$ contains a term that is
stochastic, {\em i.e.}  $\B.V\!_{_{\rm T}}$, see (\ref{wind2}).
The average of Eq.~(\ref{solPhi}) is computed in Appendix~A. The
result is:
\begin{equation}
\hat \Phi_0^{\alpha\beta}(\B.k,t)=
\Phi_0^{\alpha\beta}(\B.k) \exp\big \{ i\B.k \cdot \overline{\B.V\! _0 } \,t
-\gamma(k)\,t -2(v_{_{\rm T}}k\,t)^2  \big \}\ .
\label{hatPhi}
\end{equation}
The first term in the exponent stems from the advection by
the mean wind $\overline{\B.V\! _0 }$. The second one is the
correlation decay due to the finite life-time of the fluctuations.  The
last term in the exponent describes the effect of decorrelation due to the
random sweeping by the random component $\B.V\!_{_{\rm T}}$.

Using Eq. (\ref{compF}) we compute
\begin{eqnarray}
F^{\alpha\beta}_0(\B.R,t)&=&\int{d\B.k\over
4\pi^3}\Phi_0^{\alpha\beta}(\B.k)\Big\{1-
\exp\Big[-2(v_{_{\rm T}}kt)^2-\gamma(k)|t|
\Big]\nonumber\\&\times&
\cos(\B.k\cdot\B.R-\B.k\cdot\overline{\B.V\! _0 } \,t)\Big\} \ ,
\label{result1}
\end{eqnarray}

The structure function $S_0^{\alpha\beta}(R)$ is obtained from
(\ref{result1}) by substituting $t=0$:
\begin{equation}
S_0^{\alpha\beta}(\B.R)=\int{d\B.k\over 4\pi^3}
\Phi_0^{\alpha\beta}(\B.k)\left\{1-
\cos(\B.k\cdot\B.R)\right\} \ . \label{resultS}
\end{equation}
On the other hand $T_0^{\alpha\beta}(t)$ is obtained by putting $r=0$:
\begin{eqnarray}
T_0^{\alpha\beta}(t)&=&\int{d\B.k\over
4\pi^3}\Phi_0^{\alpha\beta}(\B.k)\Big\{1-
\exp\Big[-2(v_{_{\rm T}}kt)^2-\gamma(k)|t|\Big]\nonumber\\
&\times&\cos(\B.k\cdot\overline{\B.V\! _0 } \,t)\Big\} \ , \label{resultT}
\end{eqnarray}
We can compare the two expressions for any of the tensor
components. Since we are interested in exponents it is natural to
consider first the trace. In order to assess the sensitivity of our
results to the tensorial structure we will consider then the
longitudinal structure function:
\begin{eqnarray}\label{defS}
S_0(R)&=& \sum _{\alpha,\beta}S_0^{\alpha\beta}(\B.R)\,, \quad
T_0(t)= \sum _{\alpha,\beta}T_0^{\alpha\beta}(t)\,, \\ \nonumber
S_0^{\ell\ell}(\B.R)&=& \sum _{\alpha,\beta} S_0^{\alpha\beta}
(\B.R){R_\alpha R_\beta \over R^2}\,, \\ \nn
  T_0^{\ell\ell}(t)
&=& \sum _{\alpha,\beta} T_0^{\alpha\beta}(t){\overline{V\! _0 }
^\alpha \overline{V\! _0 }^\beta \over \overline{V\! _0 }^2}\ .
\end{eqnarray}
 Computing the trace, longitudinal projections and performing the
 angular integrations we end up with
\begin{eqnarray}\label{Sfin}
S_0(\B.R)&=& \int_0^\infty {k^2dk\over\pi^2}\Phi_0(k)\left\{1-
\Psi_0(kr)\right\} \ , \\ \nonumber
S_0^{\ell\ell}(\B.R)&=& \int_0^\infty {k^2dk\over 3 \pi^2}
\Phi_0(k)\left\{1-
\Psi_0^{\ell\ell}(kr)\right\} \ , \\ \label{Tfin}
T_0(t)&=&\int_0^\infty {k^2dk\over\pi^2}\Phi_0(k)\\ \nonumber
&&\times  \Big\{1-\Psi_0(k\,\overline{V\! _0 }\,t)
\exp\Big[-2(v_{_{\rm T}}kt)^2-\gamma(k)|t|\Big]
          \Big\} \,,     \\  \label{Tll}
T_0^{\ell\ell}(t)&=&\int_0^\infty {k^2dk\over 3 \pi^2}
\Phi_0(k)\\ \nonumber
&&\times  \Big\{1-\Psi_0^{\ell\ell}(k\,\overline{V\! _0 }\,t)
\exp\Big[-2(v_{_{\rm T}}kt)^2-\gamma(k)|t|\Big]
          \Big\} \,, \\  \label{Psi}
\Psi _0(x)&=&{\sin(x)\over x}\,, \quad
\Psi _0^{\ell\ell}(x)= 3\Big[{\sin(x)\over x^3}
-{\cos(x)\over x^2}\Big]\ .
\end{eqnarray}
Equation~(\ref{vT2}) allows one to express $v_{_{\rm T}}$ in terms of
$\Phi_0(k) $:
\FL
\begin{equation}\label{vt}
v_{_{\rm T}}^2 =\case{1}{3}\langle |\B.u_0(\B.r)|^2\rangle
=\int\!\frac{d\B.k}{12\pi^3} \Phi_0(k)
=\int\limits _0^\infty  \frac{k^2 dk}{3\pi^2}\Phi_0(k).
\end{equation}

\subsection{Assessment of the Taylor hypothesis for homogeneous advection}
The comparison between $S_0(\B.R)$ and $T_0(t)$ is determined
by the two free coefficients in this model, $C$ of Eq.~(\ref{defgam})
and
\begin{equation}
q\equiv v_{_{\rm T}}/\overline{V\! _0 } \ . \label{q}
\end{equation}
In comparing the two function we have freedom in defining the
effective advecting mean wind $V_{\rm ad}$. In the Taylor hypothesis
$V_{\rm ad}=\overline{V\! _0 }$, and one is supposed to identify
$T_0(t=|\B.R/\overline{V\! _0 }|)$ with
$S_0(\B.R)$. In some applications, when $\overline{V\! _0 }=0$ the Taylor
hypothesis
has been used \cite{91Lvo} with $V_{\rm ad}=v_{_{\rm T}}$.  In our
comparison we find it advantageous to employ an interpolation formula
\begin{equation}
V_{\rm ad} =\sqrt{\overline{V\! _0 }^2+(bv_{_{\rm T}})^2} \ , \label{defV*}
\end{equation}
with $b$ chosen to minimize the difference between the two functions
Eqs.  (\ref{Sfin},\ref{Tfin}).  Of course, for one probe measurement
the apparent {\em scaling exponent} is always independent of the
choice of the effective advective wind and of the parameter $b$ in
particular. For two or several probe measurements, when we face a
mixture of temporal and spatial contributions to the total separation
the choice of $V_{\rm ad}$ and of the parameter $b$ become important as
discussed below.
\end{multicols}{2}
\widetext
\begin{figure}
\epsfxsize=18truecm
\epsfbox{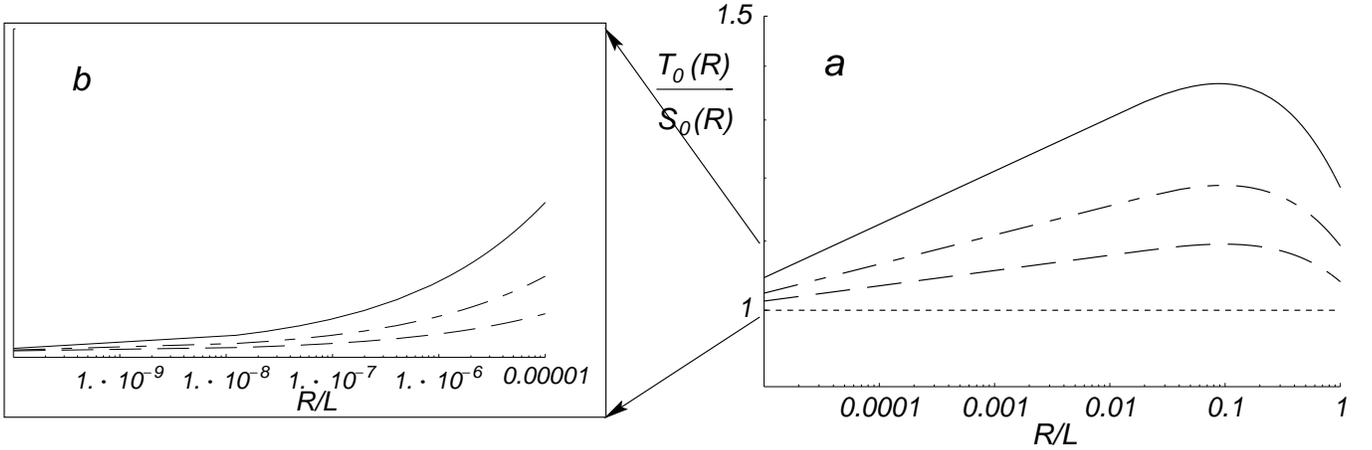}

\vskip 0.5cm
\caption{A  log-log plot of the ratio of
$T_0(R/V_{\rm ad})/S_0(R)$ vs $R/L$ for three values of $C$,
$C=$0.25 (dashed line) 0.5 (dot-dashed line) and 1
(solid line)  and $\overline{V}_0=0$ ($q\to \infty$). Panel a corresponds to
$R/L$ between 1 and $10^{-5}$, the blow-up in Panel b shows
the next five decades of $R/L$ between $10^{-5}$  and $10^{-10}$.}
\label{fig1}
\end{figure}
\begin{multicols}{2}
\narrowtext
In Fig.~\ref{fig1} we present a log-log plot of the ratio of
$T_0(R/V_{\rm ad})/S_0(R)$ vs $R/L$ for three values of $C$, $C=$0.25, 0.5
and 1, and $\overline{V}_0=0$ ($q\to \infty$).  If the Taylor hypothesis were
exact, this ratio would have been unity for all $R$.  We find that in
the limit $R/L\to 0$ the ratio of these two functions goes to a
constant which depends on the choice of $b$ in Eq. (\ref{defV*}). This
reflects the correctness of the Taylor hypothesis for $R/L\to 0$ which
follows from the fact that the sweeping time $R/V_{\rm ad}$ is negligible
compared to the life time $\propto R^{2/3}$.  The relation between the
units of distance and the units of time needs to be determined. We fix
the parameter $b$ by the requirement that $T_0(R/V_{\rm ad})$ should equal
$S_0(R)$ when $R/L\to 0$.  We found that  the effective
wind may be  approximated by Eq.~(\ref{defV*}) with

\begin{center}
\fbox{$  b \approx 3.1 $ for the modulo structure function~$S_0(R)$}\ .
\end{center}

\noindent
This fixing of the units will be of crucial importance
when we discuss two-probe measurements below.

We see that the ratio $T_0(R/V_{\rm ad})/S_0(R)$ does not scale
with $R$ when many decades of $R$ are available. In most experiments
the range of available $R$ is much smaller, and {\em apparent} scaling
will result. To demonstrate this we present in Fig.~\ref{fig1}b log-log plots
of the ratio of $T_0(R/V_{\rm ad})/S_0(R)$ vs $R/L$ for the same
values of $C$ but for $R$ values spanning only the last 5 decades of
scales. Clearly, the plots seem linear over at least 4 decades.

\begin{figure}
\epsfxsize=9truecm
\epsfbox{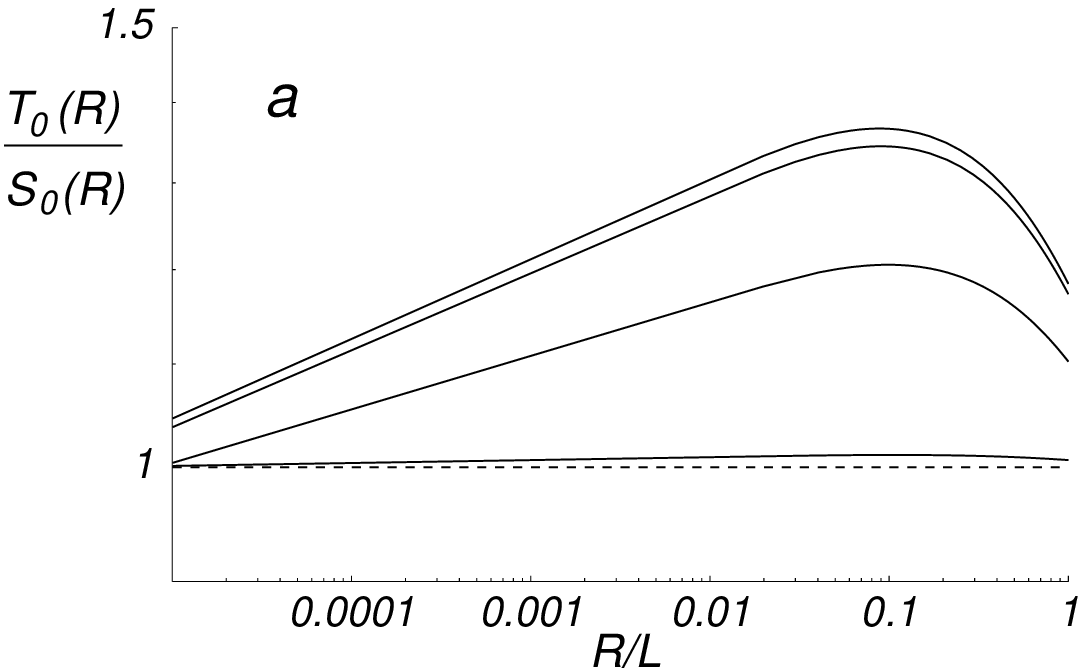}
\epsfxsize=9truecm
\epsfbox{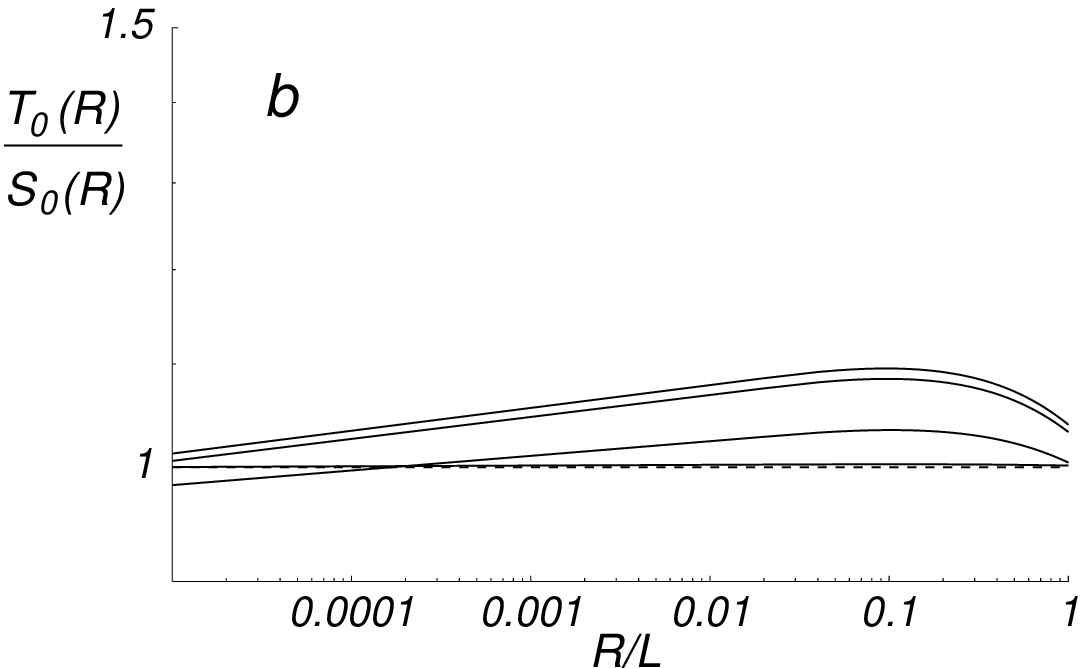}
\vskip 0.5cm
\caption{A  log-log plot of the ratio of
$T_0(R/V_{\rm ad})/S_0(R)$ vs $R/L$ for C=1 (Panel a) and C=0.25
(Panels b).  Different solid lines
correspond to values $q=10$ (the upper line), $q=1, 0.25$
{from top to bottom} and $q=0.01$ the bottom solid line. Dashed line
shows the limit $q\to 0$, when the Taylor hypothesis is exact.}
\label{fig2}
\end{figure}

In Fig.~\ref{fig2} we show log-log plots of the same ratio, for
$c=0.25$ and $C=1$, and for values of $q$ ranging from 0.01 to 10.  We
see that for $C=1$ when the mean wind is 4 times larger than $v_{_{\rm
T}}$ we have up to $20 \%$ deviations in the magnitude of
$T_0(R/V_{\rm ad})/S_0(R)$ from unity. For $q$ large (the graphs
almost saturate for $q=10$) the deviations reach 35$\%$. In terms of
the apparent scaling exponent the almost linear log-log plots can
easily deceive even an experienced researcher to conclude that the
value of $\zeta$
\noindent
 is larger than what could be measured from spatial
differences via $S_0(R)$. This finding is in agreement with
\begin{figure}
\epsfxsize=9truecm
\epsfbox{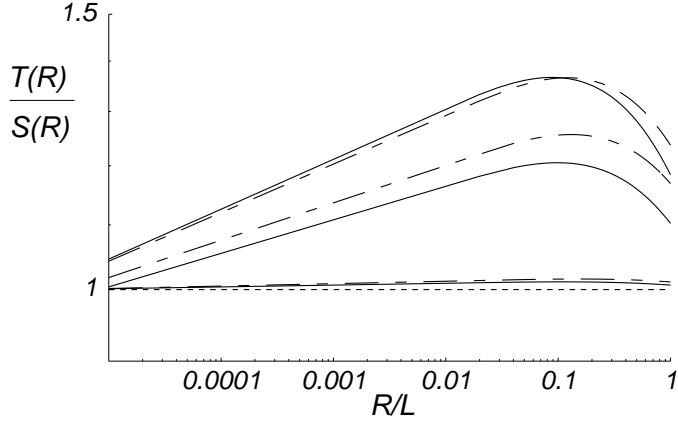}
\vskip 0.5cm   \narrowtext
\caption{A  log-log plot of the ratio of
$T_0(R/V_{\rm ad})/S_0(R)$ vs $R/L$ (solid lines) and
$T^{\ell\ell}(R/V_{\rm ad})/S^{\ell\ell}_2(R)$
 (dot-dashed lines) vs $R/L$
for for C=1.  Different lines correspond to ( from the top to the
bottom) $q=\infty $ , $q=0.25$ and $q=0.01$.}
\label{fig3}
\end{figure}
\noindent 
the
conclusion of Sreenivasan's group \cite{98DS,93Zub} who studied this
issue experimentally. Within our model we can see that the apparent
scaling exponent depends on the parameter $C$ which govern the decay
time of fluctuations, cf. Eq. (\ref{defgam}). For $C=1$ we find an
increase in the apparent exponent $\zeta_2$ between 0.01 and 0.03
depending on the value of $d$, varying from 0.1 to $\infty$. For
$C=0.25$ the increase is depressed by a factor of three. The lesson is
that for experimental applications it is very advisable to achieve a
good estimate of the inherent decay time of fluctuations of size $R$.

In order to check how our results depend on the tensorial structure of the
correlation functions we repeated the same comparisons for the longitudinal
structure functions $S_0^{\ell\ell}$ and  $T_0^{\ell\ell}$. We found
that the unit fixing parameter $b$ in this case differs from the previous one:
\begin{center}
\fbox{$  b \approx 4.2 $ for the longitudinal  structure function
$S_0^{\ell\ell}(R)$}~.
\end{center}
In order to demonstrate that apparent corrections to the scaling
exponents are similar for different tensorial components we plotted in
Fig.~\ref{fig3} the ratios $T_0(R/V_{\rm ad})/S_0(R)$ (solid lines) and
$T_0^{\ell\ell}(R/V_{\rm ad})/S^{\ell\ell}_0(R)$ (dot-dashed lines)
{\sl vs} $R$ for several values of $C$ and $q$. One sees that with the proper
choice of $b$, these ratios practically coincide.

The conclusions of this part of the analysis are as follows:
\begin{enumerate}
\item The best values of $b$
are significantly larger than the naive choice $\sqrt{3}$. They
depend on the choice of tensorial components of the correlation functions.
\item The parameter $C$, which determines the life time
$\gamma(k)$, should be
known in order to assess the systematic errors involved in Taylor
hypothesis.
\end{enumerate}
\section{The case of shear}
\subsection{Solution for linear shear}
In this section we seek the first order corrections to the 2nd order
correlation functions $S$ and $T$ which are caused by the existence of
a small shear, $U_{\rm
s}\ll \overline{V\! _0 }$. To this aim we split the
velocity field into homogeneous and shear-induced contributions:
\begin{equation}\label{u11}
\tilde \B.u(\B.k,\o)=\tilde \B.u_0(\B.k,\o)+\tilde \B.u_{\rm s}
(\B.k,\o)\,,
\end{equation}
where, as before $\tilde \B.u_0(\B.k,\o)$ is the solution with zero-shear
given by~(\ref{u0}), and $\tilde \B.u_{\rm s} (\B.k,\o)$ is induced by
the shear $V_{\rm s}$. To find $\tilde \B.u_{\rm s}$, we use
Eq.~(\ref{model}) with $\tilde \B.u(\B.k,\o)$ from (\ref{u11}),
$\tilde \B.u_0(\B.k,\o) $ from (\ref{u0}), $\B.V(\B.k)$
from~(\ref{wind1}) and (\ref{choiceV}) to get
\begin{eqnarray}
\label{u1}
\tilde\B.u_{\rm s}(\B.k,\o)&=&\tilde\B.u_\B.q(\B.k,\o)
-\tilde\B.u_{-\B.q}(\B.k,\o)       \,,  \\ \label{uq}
\tilde\B.u_{\pm \B.q}^\alpha(\B.k,\o)&=&  \case{1}{2}
V_{\rm s}  P^{\alpha\beta}(\B.k)\big[
(\B.k\cdot\B.n)\delta_{\beta\gamma}
+n^\beta q^\gamma \big]   \\ \nonumber
&\times& G_0(\B.k,\omega)  G_0 (\B.k\mp \B.q ,\omega)
\tilde f^\gamma (\B.k\mp \B.q) \ .
\end{eqnarray}
Having defined the velocity field we return to the correlation
function Eq. (\ref{Phi}) and split $\tilde\Phi^ {\alpha\beta}
(\B.k,\B.k',\o)$ into isotropic and anisotropic, shear-induced,
contributions:
\begin{equation}
\tilde\B.\Phi^{\alpha\beta}(\B.k,\B.k',\o)=(2\pi)^3
\delta(\B.k-\B.k')\tilde \B.\Phi_0^{\alpha\beta}(\B.k,\o)
+\tilde\B.\Phi_{\rm s} ^{\alpha\beta}(\B.k,\B.k',\o) \ .
\end{equation}
Here  $\tilde\Phi_0^{\alpha\beta}(\B.k,\o)$ is given
by~(\ref{Phi-ko}).  According to Eqs.~(\ref{u11}), (\ref{uq}) and
the definition (\ref{Phi}) equation for $ \tilde\B.\Phi_{\rm s}
^{\alpha\beta}(\B.k,\B.k',\o) $ may be presented as a sum:
\begin{eqnarray}\label{Phis}
\tilde\B.\Phi_{\rm s}
^{\alpha\beta}(\B.k,\B.k',\o) &=&\tilde\B.\Phi_\B.q
^{\alpha\beta}(\B.k,\B.k',\o) - \tilde\B.\Phi_{-\B.q}
^{\alpha\beta}(\B.k,\B.k',\o) \\ \nonumber
&+ &\tilde\B.\Phi_\B.q
^{*\beta\alpha}(\B.k',\B.k,\o) - \tilde\B.\Phi_{-\B.q}
^{*\beta\alpha }(\B.k',\B.k,\o)\,,
\end{eqnarray}
where
\begin{eqnarray}\label{Phiqo}
&&\tilde\B.\Phi_\B.q
^{\alpha\beta}(\B.k,\B.k',\o) =
(2\pi)^3 \delta(\B.k-\B.q-\B.k')V_{\rm s}
G_0(\B.k,\o) \\ \nonumber
&\times &P^{\alpha\delta}(\B.k)\big[
(\B.k\cdot\B.n)\delta_{\delta\gamma}
+n^\delta  q^\gamma \big]   \mbox{Im}\,\{G_0(\B.k',\o)\}
\Phi_0^{\gamma \beta}(\B.k')\ .
\end{eqnarray}
In $\B.k,t$-representation the last equations takes the form:
\begin{eqnarray}\label{Phiqt}
&&\hat \B.\Phi_\B.q
^{\alpha\beta}(\B.k,\B.k',t) \approx
(2\pi)^3 \delta(\B.k-\B.q-\B.k')\frac{V_{\rm s}}{4i\gamma_+}
P^{\alpha\delta}(\B.k)\\ \nonumber
&\times&  \big[
(\B.k\cdot\B.n)\delta_{\delta\gamma}
+n^\delta  q^\gamma \big] \Phi_0^{\gamma \beta}(\B.k')
\exp\big[(i\B.k_+\cdot \B.V\!_0
-\gamma_+)t\big]\,,
\end{eqnarray}
where we introduced
\BE\label{defs}
\B.k_+=\case{1}{2}(\B.k+\B.k')\,,\quad \gamma(k_+)=\gamma_+\ .
\EE
Having in mind the approximation of the linear shear we keep in
$\tilde\B.\Phi_{\rm s}
^{\alpha\beta}(\B.k,\B.k',\o)$ only terms that are either $q$-independent
or linear in $q$.
Correspondingly we may present~(\ref{Phiqt}) as:
\begin{eqnarray}\nonumber
&& \hat \B.\Phi_\B.q
^{ \alpha\beta}(\B.k,\B.k',t) \approx
\frac{\pi^3 V_{\rm s}}{i\gamma_+}\delta(\B.k-\B.q-\B.k')
\exp\!\big[(i\B.k_+\!\cdot \!\B.V\!_0 -\gamma_+)t\big]\\ \label{res1}
  &&\times \Big\{   P^{\alpha\beta}(\B.k_+)
  \Big[ 2\B.k_+\!\!\cdot\!\B.n  +\B.q\!\cdot\!\B.n
+ (\B.k_ +\!\! \cdot\! \B.n) (\B.q\!\cdot\!\B.k)
   \frac{\partial }{ k_+ \partial k_+}\Big]     \\ \nn
&&+2 P^{\alpha\gamma }(\B.k_+)  n^\gamma q^\delta
 P_0^{\delta  \beta}(\B.k_+)  + (\B.k_+ \!\!  \cdot \! \B.n) \frac{q^\alpha
k_+^\beta
 - q^\beta  k_+^\alpha   }{k_+^2}     \Big\} \Phi_0(k_+) \ .
\end{eqnarray}
To compute $\C.F^{\a\b}_{\rm s}(\B.R,\B.R',t)$ we need to use
the Fourier transform~(\ref{Frt-a}), which involves the integrations
$d\B.k \,d\B.k'=d\B.k_+d(\B.k-\B.k')$ and $\exp[i(\B.k\cdot \B.R-
\B.k'\cdot \B.R']$. The latter may be presented as
$$\exp[i(\B.k\cdot \B.R- \B.k'\cdot \B.R')]=\exp(i\B.k_+\cdot \B.R)
\exp[i(\B.k-\B.k')\cdot \B.r_0]\ .
$$ Here $\B.R=\B.R-\B.R'$ is the separation between probes and
$\B.r_0=\case{1}{2}(\B.R+\B.R')$ is a mean position of the probes.
Now it is customary to introduce a mixed
$(\B.k_+,\B.r_0,t)$-representation in which one integrates with
respect of $(\B.k-\B.k')$ only:
\BEA \label{krt}
\hat \C.F^{\a\b}_q(\B.k_+,\B.r_0,t)&=&
\int\frac{d(\B.k-\B.k')}{(2\pi)^3}
\hat \B.\Phi_\B.q ^{\alpha\beta}(\B.k,\B.k',t)\\ \nn
&&\times \exp[i(\B.k-\B.k')\cdot \B.r_0]\ .
\EEA
Together with Eqs.~(\ref{Phis}) and ~(\ref{res1}) this gives:
\begin{eqnarray}\label{res2}
&& \hat \C.F_{\rm s}
^{ \alpha\beta}(\B.k,\B.r_0,t) =
\frac{1}{2 \gamma(k)}
\exp\!\big\{[i\B.k\!\cdot \!\B.V\!_0 -\gamma(k)]t\big\}\\ \nonumber
  &&\times \Big\{   P^{\alpha\beta}(\B.k)
  \Big[ 2\B.k \cdot\!\B.V_{\rm s}(\B.r_0)
+  \frac {\partial V_{\rm s}^\g  (\B.r_0)}{\partial r_0^\d}
   \frac{k ^\g k ^\d  \partial }{ k\partial k}\Big]     \\ \nn
&&+ P^{\alpha\gamma }(\B.k)
\Big[  \frac {\partial V_{\rm s}^\g  (\B.r_0)}{\partial r^\d}+
 \frac {\partial V_{\rm s}^\d  (\B.r_0)}{\partial r^\g}
       \Big]
 P_0^{\delta  \beta}(\B.k)      \Big\} \Phi_0(k) \,,
\end{eqnarray}
where we redefined $\B.k_+\to \B.k$  and used the explicit
form~(\ref{prof1}) of $\B.V_{\rm s}(\B.r_0) $.

The solution~(\ref{res2}) contains a term which is proportional to the
value of the shear $\B.k\cdot\B.V_{\rm s}(\B.r_0)$ computed at the position
$\B.r_0$ between the two probes. This is just a first order term,
representing the first correction to the homogeneous velocity $\B.V_0$
due to the sweeping effect. If we were to compute higher order
sweeping corrections and were to sum them all up, we would find a
renormalized sweeping velocity in the exponent: $\B.V_0\to
\B.V_0+ \B.V_{\rm s}(\B.r_0)$. Thus instead of~(\ref{res2}) one writes:
\begin{eqnarray}\nonumber
&& \hat \C.F_{\rm s} ^{ \alpha\beta}(\B.k,\B.r_0,t) =
\frac{1}{2 \gamma(k)}
\exp\!\big\{i\B.k\!\cdot \![ \B.V_0+ \B.V_{\rm s}(\B.r_0)]t
-\gamma(k)t\big\}
\\ \label{res3}  &&\times
\Big[  \frac {\partial V_{\rm s}^\g  (\B.r_0)}{\partial r^\d}+
 \frac {\partial V_{\rm s}
^\d  (\B.r_0)}{\partial r^\g}
       \Big] \Big[   P^{\alpha\beta}(\B.k)
  k ^\g k ^\d
   \frac{ \partial \Phi_0(k)}{ \partial ^2k}    \\ \nn
&&+P^{\alpha\gamma }(\B.k) P^{\delta  \beta}(\B.k)
\Phi_0(k)    \Big] \ .
\end{eqnarray}
We should comment at this point that the calculation resulted in
an intuitively pleasing rule: effective Taylor wind should
be taken as the mean wind at the point midway between the two probes.
Also, we see that the magnitude of the shear induced part is proportional
to the shear midway between the probes. Of course, this simple rule
is a result of the assumption of {\em linear} shear. Nevertheless, as long
as the shear profile is not too nonlinear on the scale of the separation
between the two probes, this simple rule can be taken as a rule of thumb
for experimental applications.

Finally, we remember that the space homogeneous part of the wind
$\B.V_0$ has a fluctuating component, $\B.V_0 =\overline{\B.V\! _0 } +
\B.V\!_{_{\rm T}}$.  One has to average therefore the
result using the Gaussian distribution $\C.P(\B.V\!_{_{\rm T}})$.  The
final answer in analogy with~(\ref{hatPhi}) reads:
\begin{eqnarray}\label{res4}
&& \hat \C.F_{\rm s} ^{ \alpha\beta}(\B.k,\B.r_0,t)
=\C.F_{\rm s} ^{ \alpha\beta}(\B.k,\B.r_0)  \\ \nn
&\times&
\exp\!\big\{i\B.k\!\cdot \! [\overline{\B.V\! _0 }
+\B.V_{\rm s}(\B.r_0)]t-\gamma(k)t
-2(v\!_{_{\rm T}} k t)^2\big\}\,,           \\  \nn
&& \C.F_{\rm s} ^{ \alpha\beta}(\B.k,\B.r_0) =
\frac{\omega_{\rm s} (\B.r_0)       }{2 \gamma(k)}
\Big[  2 P^{\alpha\beta}(\B.k)(\B.k\cdot \B.n)(\B.k\cdot \B.m)
 \frac{ \partial \Phi_0(k)}{ \partial ^2k}    \\ \nn
&&+ P^{\alpha\gamma }(\B.k) (n^\g m^\d + n^\d m^\g )
 P^{\delta  \beta}(\B.k)
\Phi_0(k)    \Big] \,,
\end{eqnarray}
were in agreement with~(\ref{prof1}) we introduced a ``shear frequency''
$\omega_{\rm s}(\B.r)$ according to
\BE\label{den1}
\frac {\partial V_{\rm s}^\a  (\B.r_0)}{\partial r^\b}\equiv
\omega_{\rm s}  (\B.r_0)  n^\a m^\b\ .
\EE
Examining Eq. (\ref{res4}) we see that the scaling exponent expected
for $\C.F_{\rm s} ^{ \alpha\beta}(\B.k,\B.r_0)$ is determined by the
scaling of $\Phi_0(\B.k)$ and $\gamma(k)$, With the choices specified
in Eqs. (\ref{defgam}) and (\ref{choice}) $\C.F_{\rm s} ^{
\alpha\beta}(\B.k,\B.r_0)
\propto k^{-13/3}$, or $R^{4/3}$ for the second order structure function.
This is consistent with the expected scaling in the anisotropic sector
characterized by $j=2$, and see \cite{99ALPb} for more details.

Note that  in the case linear shear the frequency
$\omega_{\rm s} (\B.r) $ is $\B.r$
independent.  Similarly to Eqs.~(\ref{resultS}) and (\ref{resultT})
one computes the shear induced additions of $S^{\a\b}_{\rm s} (\B.R)$ and
$T^{\a\b}_{\rm s} (t)$ to the usual and Taylor-computed structure functions
$S^{\a\b}(\B.R)$ and $T^{\a\b} (t)$:
\BEA
S_{\rm s} ^{\alpha\beta}(\B.R)&=&\int{d\B.k\over 4\pi^3}
\C.F_{\rm s} ^{\alpha\beta}(\B.k)\left\{1-
\cos(\B.k\cdot\B.R)\right\} \,,\label{resultsST}\\
T_{\rm s} ^{\alpha\beta}(t)&=&\int{d\B.k\over
4\pi^3}\C.F_{\rm s} ^{\alpha\beta}(\B.k)\Big\{1-
\exp\Big[-2(v_{_{\rm T}}kt)^2-\gamma(k)|t|\Big]\nonumber\\
&\times&\cos\big\{\B.k\cdot[\overline{\B.V\! _0 }
+ \B.V_{\rm s}(\B.r_0) \big]\,t\big\}\Big\} \ , \nn
\end{eqnarray}
In experimental measurements we can isolate the shear induced contribution
on the expense of the isotropic
contribution by considering a mixed, transverse-longitudinal
structure function, taking the separation $\B.R$ along the wind
$\B.R_\ell=\B.n (\B.R\cdot \B.n)$.  For example:
\BE\label{mix1}
S_{\rm s} ^{t\ell}(R) \equiv
S_{\rm s} ^{\alpha\beta}(\B.R_\ell) m^\a n^\b\,, \quad
T_{\rm s} ^{t\ell}(t) \equiv
T_{\rm s} ^{\alpha\beta}(t) m^\a n^\b\ .
\EE
These functions may be obtained from equations similar to~(\ref{resultsST})
with the replacement
\BEA\label{den2}
&& \C.F_{\rm s} ^{\alpha\beta}(\B.k)\to \C.F_{\rm s}^{t\ell}
\equiv  \C.F_{\rm s} ^{\alpha\beta}(\B.k)m^\a n ^\b \\ \nn
&=&\frac{\omega_{\rm s} }{\g(k)}\Big\{ - \frac{(\B.k\cdot \B.n)^2
(\B.k\cdot \B.m)^2}{k^2}\frac{d\, \Phi_0(k)}{d\,k^2}\\
&+&\frac{1}{2}\Big[1- \frac{(\B.k\cdot \B.n)^2}{k^2}\Big]
\Big[1- \frac{(\B.k\cdot \B.m)^2}{k^2}\Big]
\Phi_0(k)\Big\}\ .
\EEA
Integration this over $\phi$, the azimuthal angle of $\B.k$ around
direction of $\B.n$, one has
\BEA\label{den3}
\int\limits_0^{2\pi} \d\phi  \C.F_{\rm s}^{t\ell}&=&
\frac{\pi \omega_{\rm s} \sin^2 \theta }{\g(k)}
\Big\{-2 \frac{d\, \Phi_0(k)}{d\,k^2}\cos^2\theta
\\ \nn
&&  +
\Phi_0(k)(1+\cos^2 \theta) \Big\}\,,
\EEA
where $\cos\theta=\B.n\cdot\B.k/k$.  Having this in mind and
performing in~(\ref{resultsST}) the $\theta$-integration we end up
with
\begin{eqnarray}\label{Slt}
&&S_{\rm s}^{t\ell } (\B.R)=  \frac{\omega_{\rm s}}{5  \pi^2}
\int\limits _0^\infty
 {k^2dk\over \gamma(k)}\Big\{ \Phi_0(k)\big[1-
\Psi_{\rm s}^{t\ell }(kr) \big] \\ \nonumber
&&\qquad\qquad   -\frac{k^2}{3}\frac{d \Phi_0(k)}{d\,k^2}
\big[1-\overline{\Psi} _{\rm s}^{t\ell }(kr) \big]\Big\} \,,
\\ \label{Tlt}
&& T_{\rm s}^{t\ell } (t)=\frac{\omega_{\rm s}}{5  \pi^2}
\int\limits _0^\infty \!{k^2dk\over \gamma(k)}
\Big\{ \Phi_0(k)\\ \nn
&&\qquad \times  \big[1-\Psi_{\rm s}^{t\ell }(kV_{\rm ad}t)
\exp\Big[\!-\!2(v_{_{\rm T}}kt)^2\! -\! \gamma(k)t\Big]
   \\  \nn
&&
-  \frac{k^2}{3}\frac{d \Phi_0(k)}{d\,k^2}
\big[1-\overline{\Psi} _{\rm s}^{t\ell }( kV_{\rm ad}t  )
\exp\Big[\!-2(v_{_{\rm T}}kt)^2\! -\gamma(k)t\Big]
\big]\Big\}
\\ \nonumber
&&\Psi _{\rm s}^{t\ell } (x)=5  \big[\frac{6-x^2}{x^4}
\cos x +  3 \frac{x^2-2}{x^5}\sin x \big]\\ \label{Psilt}
&& \qquad \simeq 1
-\frac{5x^2}{42} \,, \\ \nn
&& \overline{\Psi} _{\rm s}^{t\ell } (x)=15  \big[\frac{12 -x^2}{x^4}
\cos x + \frac{5 x^2-12}{x^5}\sin x \big]
\simeq 1 -\frac{5x^2}{14}\ .
\end{eqnarray}
Formally  expansion of  $\Psi _{\rm s}^{t\ell } (x)$  and
$\overline{\Psi} _{\rm s}^{t\ell } (x)$ at small $x$ begin with
$1/x^4$ terms, but due to double cancellation it actually starts from $1$.
We analyzed numerically Eqs.~(\ref{Slt}~--~\ref{Psilt}) in the following
Subsection.

\subsection{Discussion of the case of shear}
The first difference between Eqs.~(\ref{Slt} -- \ref{Psilt}) for the
anisotropic contribution to the structure functions $S_{\rm s}^{t\ell }$,
 $T_{\rm s}^{t\ell }$ and the corresponding structure functions
$S_0^{\a\b}$, $T_0^{\a\b}$ is in their scaling behavior. In the
integrals~(\ref{Sfin}~--~\ref{Psi})
for $S_0^{\a\b}$,   $T_0^{\a\b}$ the function
$\Phi_0(k) \propto k^{-11/3}$. These integrals converge, and the main
contribution comes from the region $kR\sim 1$. Both quantities scale
according to $S_0^{\a\b}(R)
\propto R^{2/3}$ and    $T_0^{\a\b}(R)
\propto R^{2/3}$ in the limit  $R/L\to 0 $,  as expected. In contrast to that,
the integrands in Eqs.~(\ref{Slt} -- \ref{Psilt})   have an additional factor
$\g(k)\propto k^{2/3}$ in the
denominator. This changes the scaling behavior to $S_{\rm s}^{t\ell }(R)
\propto R^{4/3}$, $T_{\rm s}^{t\ell }(R)
\propto R^{4/3}$. The second difference is in the rates of the convergence.
The integrals for $S_0^{\a\b}$ and   $T_0^{\a\b}$ behave in the region of
$kL\ll 1$ like $\int_0 k^{1/3}dk $ while the integrals for
$S_{\rm s}^{t\ell }$ and
 $T_{\rm s}^{t\ell }$ behave in the region of small $k$ like
$\int_0 k^{-1/3}dk$. One sees that the latter 
\begin{figure}
\epsfxsize=9truecm
\epsfbox{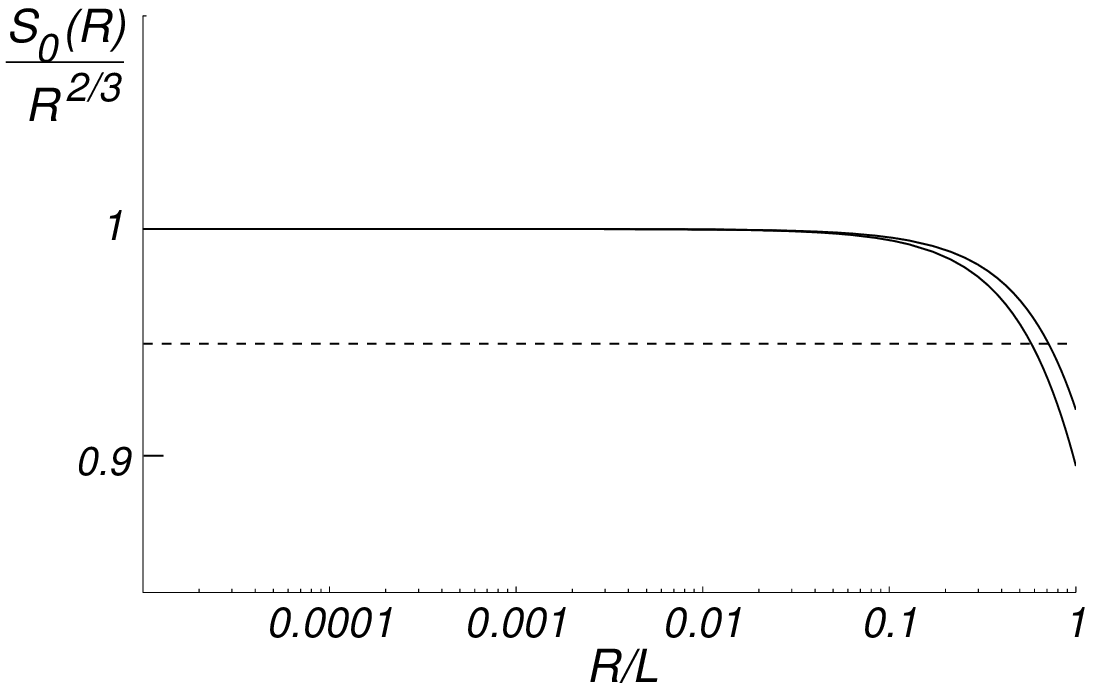}
\epsfxsize=9truecm
\epsfbox{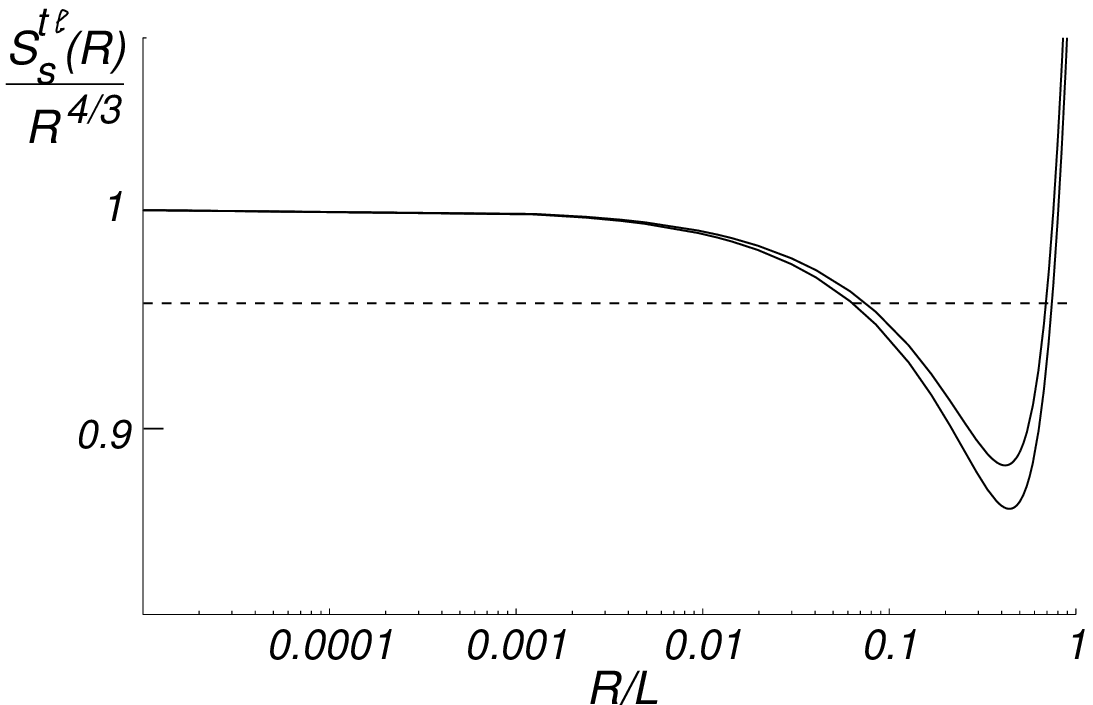}
\vskip 0.5cm   \narrowtext
\caption{A  log-log plot of the ratio of
$S_0(R)/R^{2/3}$ vs $R/L$ (the top  Panel)  and
$S^{t\ell}_{\rm s}(R)/R^{4/3}$ (the bottom Panel) vs $R/L$
 Different lines correspond to different choices (29,30) of the power spectrum
$\Phi_0(k)$. Dashed line denotes level 0.95.}
\label{fig4}
\end{figure}
\noindent
integrands have an integrable
singularity. The contribution of the nonuniversal energy containing region
$kL\sim 1$ is much more
pronounced than in the corresponding integral $\int_0 k^{1/3}dk $.
 As a consequence
one needs to consider much smaller values of $R/L$ to see the
asymptotic scaling of the functions $S_{\rm s}^{t\ell }$ and
$T_{\rm s}^{t\ell }$ compared to the case of  $S_0^{\a\b}$
and  $T_0^{\a\b}$.

This is illustrated in the
Fig.~\ref{fig4}. Panel $a$ shows log-log plots of $S_0(R)/R^{2/3}$
vs $R/L$ for two  choices  (\ref{choice}) and (\ref{choice2}) of cutoffs
functions $\Phi_0$. The two lines almost coincide and reach a
level of 0.95 at $R\simeq L/3$. Panel $b$  exhibits the corresponding
log-log plots of $S_{\rm s}^{t\ell }(R)/R^{4/3}$ vs $R/L$. The plots
reach the level 0.95 at much  smaller $R$ values (about $R\simeq L/10$),
as expected. However the two plots are significantly different only when
there is no scaling behavior (for $R>L/3$). We thus propose that our
main findings  are independent of the  choice of the crossover behavior
of the power spectrum $\Phi_0(k)$ (within reasonable choices of the 
functions $\Phi_0$). 

As mentioned above, for small mean winds
the Taylor method is problematic for large values of $R$ but
it improves for smaller values. Therefore  the significantly 
\begin{figure}
\epsfxsize=9truecm
\epsfbox{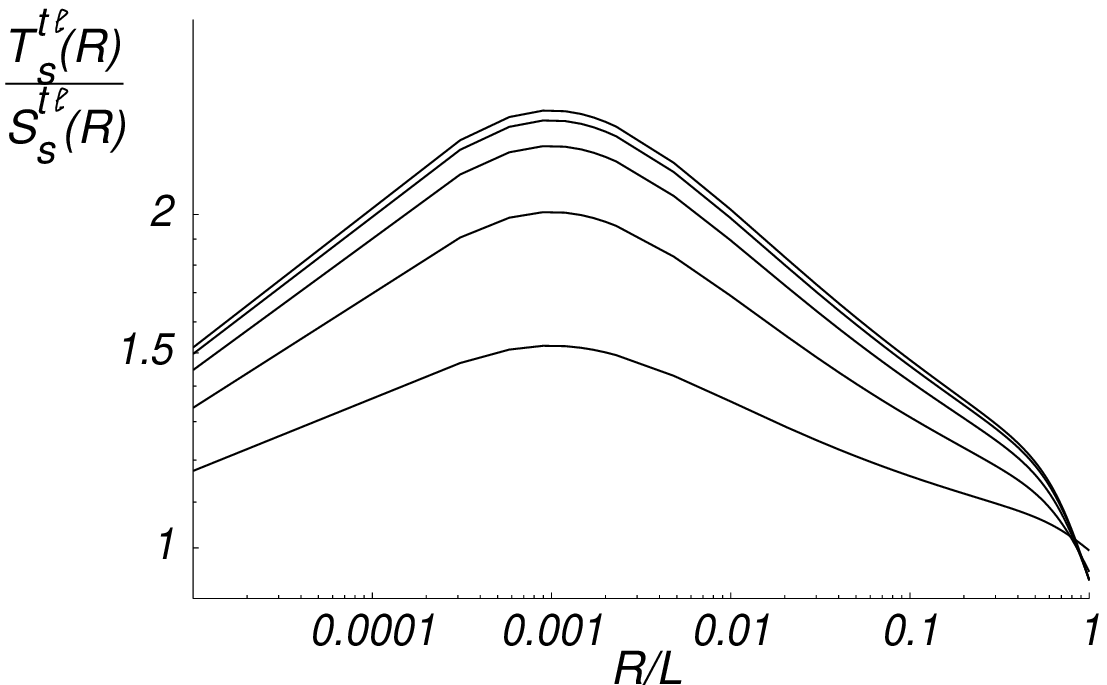}
\epsfxsize=9truecm
\epsfbox{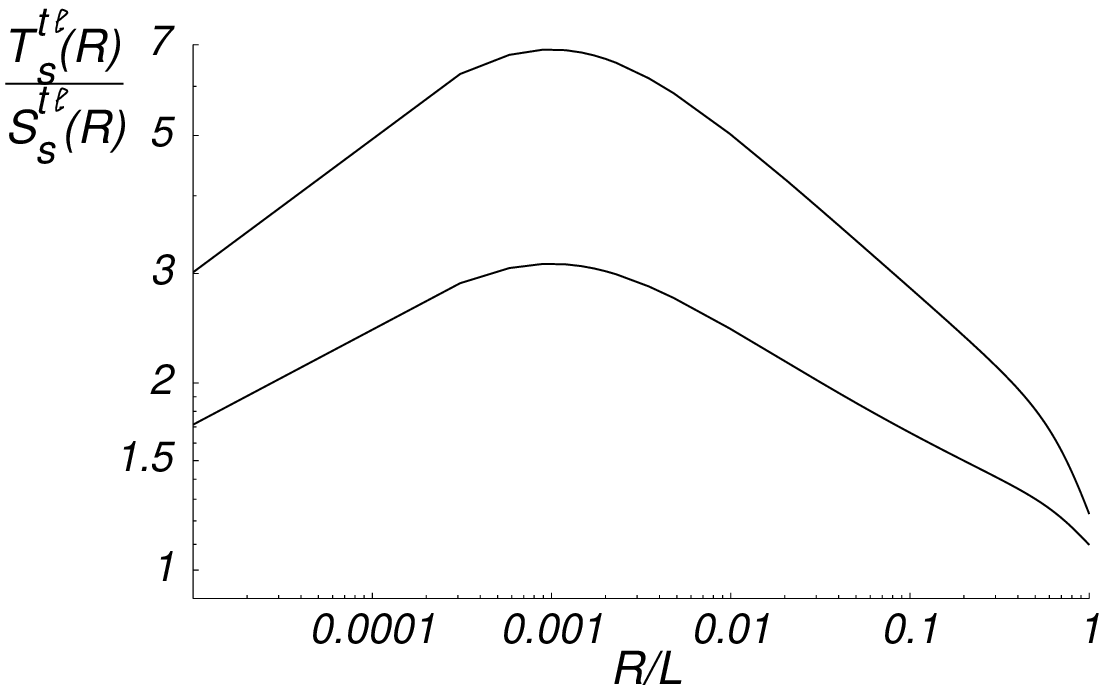}
\vskip 0.5cm
\caption{On the top Panel: A  log-log plot of the ratios
$T^{t\ell}_{\rm s}(R/V_{\rm ad})/S^{t\ell}_{\rm s}(R)$
 vs $R/L$ for  $C=0.25$  and different      values of $q=10$
(the upper line), $q=1, 0.5, 0.25$
{from top to bottom} and $q=0.1$ the bottom line.  The bottom Panel
represents the ratios  for $C=1$. The top line correspond to
$q=10$, the bottom line -- $q=0.1$.}
\label{fig5}
\end{figure}
\noindent
more
pronounced contribution of the large scale eddies for
the shear-induced part of the structure functions (in comparison with
the isotropical one)  has to lead to larger deviations of the Taylor surrogate
$T_{\rm s}^{t\ell }(R)$ from the directly measured structure function
$S_{\rm s}^{t\ell }(R)$.  This is illustrated by  the log-log plots
of the ratio $T_{\rm s}^{t\ell }(R)/S_{\rm s}^{t\ell }(R)$ vs. $R/L$
in Fig.~\ref{fig5}. 

The top  Panel  represents this ratio for $c=0.25$
and for values
of the parameter $d$ ranging between $d=0.1$ (the lower line),
$d=0.25, 0.5, 1$ and $d=\infty$  (the upper line). In contrast to
isotropic case we have here
two regimes, one with negative apparent correction to the scaling
exponent (in the
region $10^{-3}L < R < 0.3L$) and a second with a positive
correction (for $R< 10^{-3}L $). The largest possible corrections
are obtained in the absence
of the mean wind
($d=\infty$), reaching $\pm 0.13$. For $d=0.25$ the corrections
are about $\pm 0.1$
and for $d=0.1$ they are about $\pm 0.06$
Te bottom Panel  shows the ratio for $c=1$ and $d=\infty$,
(the upper line) and $d=0.1$, (the lower line). The corrections to the
apparent
scaling exponents are  $\pm 0.21 $ and $\pm 0.15 $ respectively.
The conclusion is that in the absence of the mean wind ($d=\infty$) one has
to be
weary of using the Taylor surrogate instead of direct
measurements in space. If the mean wind is relatively large (say, $d<0.1$,
as is
quite common) the expected error in the scaling exponent is about $0.1$. This
is definitely a large error but it is substantially smaller than the
difference between the isotropic and the shear induced exponents for the
second order
structure functions (2/3).

\section{Summary and discussion}
In this paper we presented an exactly soluble model of an advected
field whose fluctuations are chosen to mimic as closely as possible
those of turbulence with K41 spectra. The aim is to assess the accuracy
of the Taylor surrogate structure function by solving exactly for
the space dependent and the time dependent 2nd order structure functions
and comparing between them. Clearly, the most important consideration is the
decay time of correlations of size $R$ compared to the rate of sweeping across
$R$. The parameter $C$ in our model determines the ratio of the turnover time
to the decay time, and is free in our model.

The main results of the analysis are as follows:
\begin{enumerate}

\item For data extracted from a single probe in isotropic flows the
error introduced by the Taylor method is systematic, always leading
to an over-estimate in the scaling exponent of the 2nd order structure
function. This is in agreement with the conclusions of Sreenivsan's group
who studied this issue experimentally \cite{98DS,93Zub}.

\item The error in the isotropic scaling exponent  which is introduced by the Taylor method is typically small,
reaching 0.01 in the most adverse situation.

\item The rms velocity is an important contribution to the effective wind,
and should not be left out. Eq.(\ref{defV*}) is a simple recipe that
can be followed, with $b$ chosen to minimize the errors. We found that
our model yields the smallest errors with $b\approx 3.1$.

\item For data extracted from two probes in anisotropic fields the best rule
of thumb is to use the mean velocity and mean rms of the two probes. The best
value of $b$ for the model treated above is $b=3.8$.

\item The errors introduced by the Taylor method in anisotropic fields
are considerably larger than those found in isotropic flows. In the most
adverse situation errors in the scaling exponents can reach 0.15.
Worse, they are not systematic, tending from positive errors for smaller
scales to negative errors for larger scales.

\item Nevertheless, the errors are significantly smaller than the
difference between
the exponents in the different sectors of the symmetry group. Thus, the Taylor
approach can be used (with care) to extract the universal exponents
characterizing
the different sectors. An example of such an approach can be found in
\cite{99KLPS}.

\end{enumerate}
Even though these results are found on the basis of a simple
model, there are aspects that appear relatively model independent.
The source of error in the Taylor method is the finite life time
of the fluctuations, and the parameter $C$ that appears in the model,
the ratio of this to the sweeping time, is going to appear in
a similar fashion in any
other model or experiment. The relative improvement of the
Taylor estimates with decreasing scales is also model independent.
The need for a ``unit fixer" like $b$ is generic as well, especially
when we mix spatial and temporal distances, as is the case with
data measured by two probes. We thus hope that the analysis presented
above would be of some use for assessing experimental data as long
as the Taylor surrogates have not been replaced by direct methods
of measurements.

\acknowledgments
This
work has been supported in part by the Israel Science Foundation
administered by the Israel Academy of Sciences and Humanities,
the German-Israeli
Foundation,  The European Commission under the TMR program,
and the Naftali and Anna
Backenroth-Bronicki Fund for Research in Chaos and
Complexity.

\end{multicols}

\end{document}